\documentclass[a4paper,11pt,aps,tightenlines,nofootinbib]{revtex4}
\usepackage{xspace}
\usepackage[english]{babel}
\usepackage{amsfonts}
\usepackage{amsmath, amsthm, amssymb, mathrsfs}
\usepackage[dvips]{graphicx}
\usepackage{pdfpages}
\usepackage{setspace}
\usepackage{indentfirst}
\usepackage[left=2.5cm, right=2.5cm, top=3cm, bottom=2cm]{geometry}
\usepackage{bbm}

\usepackage{pstricks}
\usepackage{enumerate}
\usepackage{pstricks-add}
\usepackage{epsfig}
\usepackage{pst-grad}
\usepackage{pst-plot}

\usepackage{color}
\usepackage{soul}

\newcommand{\abs}[1]{\left|#1\right|}
\newcommand{\bra}[1]{\langle #1 |}

\newcommand{\ket}[1]{|#1\rangle}

\newcommand{\mean}[1]{\left\langle #1 \right\rangle}

\newcommand{\be}{\begin{equation}}
\newcommand{\ee}{\end{equation}}
\newcommand{\ba}{\begin{array}{c}}
\newcommand{\ea}{\end{array}}

\newcommand{\Cgr}{C^{\text{gr}}}
\def\beqr{\begin{eqnarray}}
\def\eeqr{\end{eqnarray}}

\numberwithin{equation}{section}
\numberwithin{thr}{section}
\numberwithin{chr}{section}
\numberwithin{df}{section}

\newcommand{\makeSymbol}[1]{\mathord{\vcenter{\hbox{#1}}}}

\newcommand{\bbra}[1]{\bigl\langle#1\bigr|}
\newcommand{\kket}[1]{\bigl|#1\bigr\rangle}
\newcommand{\bbrakket}[2]{\bigl\langle#1\big|#2\bigr\rangle}

\begin{document}

\title{Time evolution in deparametrized models of loop quantum gravity}

\author{Mehdi Assanioussi}
\email[]{mehdi.assanioussi@fuw.edu.pl}
\author{Jerzy Lewandowski}
\email[]{jerzy.lewandowski@fuw.edu.pl}
\author{Ilkka M\"{a}kinen,}
\email[]{ilkka.makinen@fuw.edu.pl}
\affiliation{Faculty of Physics, University of Warsaw, Pasteura 5, 02-093 Warsaw, Poland\\}

\begin{abstract}
An important aspect in understanding the dynamics in the context of deparametrized models of LQG is to obtain a sufficient control on the quantum evolution generated by a given Hamiltonian operator. More specifically, we need to be able to compute the evolution of relevant physical states and observables with a relatively good precision. In this article, we introduce an approximation method to deal with the physical Hamiltonian operators in deparametrized LQG models, and apply it to models in which a free Klein-Gordon scalar field or a non-rotational dust field is taken as the physical time variable. This method is based on using standard time-independent perturbation theory of quantum mechanics to define a perturbative expansion of the Hamiltonian operator, the small perturbation parameter being determined by the Barbero-Immirzi parameter $\beta$. This method allows us to define an approximate spectral decomposition of the Hamiltonian operators and hence to compute the evolution over a certain time interval. As a specific example, we analyze the evolution of expectation values of the volume and curvature operators starting with certain physical initial states, using both the perturbative method and a straightforward expansion of the expectation value in powers of the time variable. This work represents a first step towards achieving the goal of understanding and controlling the new dynamics developed in \cite{AALM, LQGSC}.
\end{abstract}

\maketitle

The Hamiltonian formulation of general relativity encodes the dynamics in constraints. This translates in a frozen picture of the dynamics where there is no time flow nor evolution of physical quantities. This specific aspect raises several serious issues in the interpretation of a quantum theory of gravity, as one fails to make the link to the experimental setup with definite instants of time. One of the approaches to circumvent this problem of time is the deparametrization of gravity \cite{KucTor1,KucTor2,RovSmo94,KucharRomano,BroKuc,BicKuc,Mark,Koul,HusPaw,GieThiem12,Swiez}. This approach however carries the drawback of choosing a specific global reference frame to parametrize either time or both space and time, hence the description and interpretation of the physics derived within the framework would be tied to this choice of the frame. Nevertheless, this approach turns out to be technically very efficient in constructing complete quantum models where gravity is fully quantized \cite{DGKL,GieThiem12,AALM}, bypassing the difficulties encountered in the case of the standard vacuum theory. Those models then become a very rich playground to test the many technical steps of the quantization procedures along with the development and analysis of new methods and ideas to answer even more complex questions concerning the semi-classical and coherent states, the quantum observables and the continuum limit of the quantum theory.

In this article we consider two LQG \cite{lqgcan1, lqgcan2, lqgcan3, lqgcan4} models where deparametrization of the scalar constraint is performed with respect to the free Klein-Gordon scalar field \cite{RovSmo94,DGKL,GieThiem12,AALM} and the non-rotational dust field \cite{HusPaw,GieThiem12,Swiez}. The main difference between the two models is the final form of the physical Hamiltonian which dictates the dynamics of the gravitational degrees of freedom with respect to the relational time provided by the considered scalar field. Following the quantization presented in \cite{AALM, LQGSC}, the Hamiltonian operators in the quantum theories are densely defined on a non-separable physical Hilbert space. Since a complete spectral decomposition of those Hamiltonian operators is so far unavailable, it is imperative to develop and use approximate methods in the analysis of the dynamics induced by these operators. A straightforward approach is to consider the expansion of the evolution operator in powers of the time parameter and introduce a truncation of the expansion at a certain fixed order of time. In this case the evolution operator reduces to a finite sum of terms where each one is calculated through a finite number of successive actions of the Hamiltonian operator. Such truncation forms a valid approximation of the time evolution when the time interval under consideration is {\it sufficiently} small. Nevertheless, this method is not enough to concretely calculate the evolution in the deparametrized model with a massless scalar field. An interesting alternative, which we propose here, is to use standard time-independent perturbation theory of quantum mechanics to introduce a perturbative expansion of the Hamiltonian operator, the small perturbation parameter being determined by the Barbero-Immirzi parameter $\beta$, which requires $\beta\gg 1$. This method allows us to define an approximate spectral decomposition of the Hamiltonian operators and hence to compute the evolution in a certain time interval.


\section{Deparametrized models and physical Hamiltonians}
In the following we will present a short overview of the two deparametrized models we are interested in. Then we will briefly present the LQG quantization of the two models and the complete quantum theories. Note that in both models the spatial diffeomorphism constraints are not solved classically and they will be carried to the quantum theory.

\subsection{Gravity coupled to a massless scalar field}
As any covariant theory, general relativity minimally coupled to a scalar field $\phi$ in its Hamiltonian formulation is set as a fully constrained system. Using Ashtekar-Barbero canonical variables $(A^i_a, E_j^b)$ for the gravitational field ($a,b$ are spatial indices while $i,j$ internal $SU(2)$ indices) \cite{A variables, B variables}, the Hamiltonian analysis reveals the following constraints \cite{Rov-Smo, DGKL}:
\begin{align}
G_i(x)&= \partial_a E_i^a + \epsilon_{ij}{}^k A^j_a E^a_k,\\
C_a(x)&= \Cgr_a(x) \ +\ \pi(x)\phi_{,a}(x),\label{DiffC}\\
C_{\ }(x)&= \Cgr(x)\ +\ \frac{1}{2}\frac{\pi^2(x)}{\sqrt{q(x)}} +\frac{1}{2}\sqrt{q(x)}E_i^a(x)E_i^b(x)\phi_{,a}(x)\phi_{,b}(x),\label{scalar}
\end{align}
the Gauss constraints, the spatial diffeomorphism constraints and the scalar constraints respectively. The quantity $\pi(x)$ is the conjugate momentum to $\phi(x)$, $q(x):=\abs{\epsilon_{abc}\epsilon^{ijk}E_i^a(x)E_j^b(x)E_k^c(x)}$ is the determinant of the densitized triad $E_i^a(x)$, and the functionals $C_a^{\rm gr}(x)$ and $C^{\rm gr}(x)$ in the above constraints are the gravitational parts. These have the following expressions
\begin{align}
C_a^{\rm gr}(x) &= \frac{1}{k\beta} F_{ab}^i(x) E_i^b(x),\\
C^{\rm gr}(x) &= -\frac{1}{16\pi G \beta^2} \biggl(\frac{\epsilon_{ijk}E^a_i(x)E^b_j(x)F_{ab}^k(x)}{\sqrt{|\det E(x)|}} + (1+\beta^2)\sqrt{|\det E(x)|} \,R(x)\biggr),
\end{align}
$G$ being the Newton constant, $\beta$ the Barbero-Immirzi parameter, $F_{ab}^i(x)$ the curvature of the connection $A^i_a(x)$, and $R(x)$ the Ricci scalar obtained from the metric tensor $q_{ab}$ on the $3$-dimensional Cauchy hypersurface $\Sigma$, the relation between $E^a_i$ and the inverse metric $q^{ab}$ being given by $q^{ab}(x) = \frac{1}{q(x)}E_i^a(x)E_i^b(x)$. In this article we call Euclidean part the part of the scalar constraint containing $F_{ab}^i(x)$, and Lorentzian part the part containing $R(x)$.

The deparametrization procedure sums up to rewriting the scalar constraint $C(x)$ so that it is linear in the momenta $\pi(x)$ and the explicit dependence on the field $\phi(x)$ is removed. The scalar field $\phi(x)$ can be then chosen to be the physical time. Assuming that the constraints \eqref{DiffC} are satisfied, the scalar constraints take the form
\begin{align}
C(x)= \Cgr(x)\ +\ \frac{1}{2}\frac{\pi^2(x)}{\sqrt{q(x)}} +\frac{1}{2 \pi^2(x)}\sqrt{q(x)} E_i^a(x)E_i^b(x)C_a^{\rm gr}(x)C_b^{\rm gr}(x).\label{scalar2}
\end{align}
Solving this equation for the momenta\footnote{Sign choices arise in the expression of $\pi(x)$ when solving \eqref{scalar2}. Those choices corresponds to treating different regions of the phase space:
\begin{align}
\pi^2\ \ge / \le \ q^{ab}(x)\phi_{,a}(x)\phi_{,b}(x)\,q(x)\ .
\end{align}
We choose the phase space region corresponding to
 \be \pi \ge \abs{q^{ab}(x)\phi_{,a}(x)\phi_{,b}(x)\,q(x)}\ , \text{and}\ -\sqrt{q}C^{gr}+\sqrt{q}\sqrt{({C^{gr}})^2-q^{ab}C^{gr}_aC^{gr}_b}\geq0 \label{Cond1}\ , \ee which interestingly contains the sector of spatially homogeneous spacetimes \cite{DGKL}. The second condition on the gravitational constraints must also be implemented in the quantum theory.} $\pi(x)$ leads to a new form of the scalar constraint $C'(x)$ equivalent to \eqref{scalar} in a specific region of phase space, namely
\begin{align}
C'(x)= \pi(x) - \sqrt{-\sqrt{q}C^{gr}+\sqrt{q}\sqrt{({C^{gr}})^2-q^{ab}C^{gr}_aC^{gr}_b}}=:\pi(x) - h_{SF}(x)\ .\label{newscalar}
\end{align}

The constraints $C'(x)$ strongly commute \cite{KucharRomano} and a Dirac observable $\mathscr{O}(x)$ on the phase space, i.e. a function which commutes with the new set of constraints, would satisfy
\begin{align}
 \frac{d \mathscr{O}}{d \phi(x)}=\{\mathscr{O},h_{SF}(x)\},
\end{align}
This equation shows precisely how the quantity $h_{SF}(x)$ arises as a physical Hamiltonian density in the reference frame  of the scalar field $\phi$, that it is the foliation with slices of constant value of the scalar field. Note that $h_{SF}(x)$ is a functional of the gravitational variables only, hence all the redundant degrees of freedom in the scalar constraints \eqref{newscalar} are absorbed in the scalar field $\phi$. The dynamics of the system is then promoted from imposing constraints, to describing evolution of the gravitational degrees of freedom with respect to the physical time set by the scalar field.

\subsection{Gravity coupled to non-rotational dust}

The model of gravity coupled to non-rotational dust \cite{BicKuc,Giesel:2007wn,HusPaw,GieThiem12,Swiez} is to some extent very similar to the one of gravity coupled to a massless scalar field described above. The difference arises from adding a potential term in the action of the system which is analogous to a cosmological constant term. While the Gauss and the spatial diffeomorphism constraints are identical to the massless scalar field case, the mentioned difference appears explicitly in the scalar constraints of the theory, namely
\begin{align}
C(x)&= \Cgr(x)\ +\ \frac{1}{2\rho}\frac{\pi^2(x)}{\sqrt{q(x)}} +\frac{\rho}{2}\sqrt{q(x)} E_i^a(x)E_i^b(x)\phi_{,a}(x)\phi_{,b}(x)+\frac{\rho}{2}\sqrt{q(x)},\label{scalar3}
\end{align}
where $(\phi,\pi)$ are the dust field variables and $\rho$ is a Lagrange multiplier appearing in the action of the system and which must satisfy certain second class constraints. Replacing $\rho$ in \eqref{scalar3} by its explicit form obtained from solving the second class constraints, and using the diffeomorphism constraints as in the previous case, we obtain the new simplified scalar constraints
\begin{align}
C'(x)= \pi(x) + C^{gr} =:\pi(x) - h_{D}(x).\label{newscalar2}
\end{align}
This equation, similarly to \eqref{newscalar}, presents the quantity $h_{D}(x)$ as the physical Hamiltonian density for the dynamics of the gravitational degrees of freedom in the reference frame of the dust. 

\subsection{Quantum theory}
The quantization is performed along the canonical program of LQG (see \cite{DGKL,AALM} for the scalar field case). In both models, the kinematical Hilbert space ${\cal H}_{\rm kin}$ is defined as the completion (with respect to the norm defined by a natural scalar product \cite{AL-measure}) of the space of \emph{cylindrical functions} of the connection variable $A$, i.e. functions depending on the differential $1$-form $A =\ A^i_a\tau_i\otimes dx^a$ (with $\tau_i$ the generators of the $su(2)$ algebra) through finitely many holonomies of the connection, which are $SU(2)$ group elements. The space ${\cal H}_{\rm kin}$ admits a basis called the spin network basis, where each element is labeled by a closed embedded graph, spins on the edges of the graph corresponding to $SU(2)$ representations of the holonomies, and $SU(2)$ invariant tensors at the vertices of the graph intertwining the representations meeting at those vertices. The fundamental operators are holonomies $\hat{h}^{(l)}_{\gamma}$, acting as multiplicative operators, defined with arbitrary embedded curves $\gamma$ and in arbitrary $SU(2)$ irreducible representations $l$, and derivative operators $\hat{J}_{x,\gamma,i}$ associated to curves $\gamma$ starting at a point $x$ in $\Sigma$ and acting in the $su(2)$ algebra.

The Gauss and spatial diffeomorphism constraints are then implemented through a group averaging procedure \cite{Ashtekar:1995zh}. The resulting space is a Hilbert space of $SU(2)$ gauge invariant and spatial diffeomorphism invariant states, we denote it ${\cal H}^G_{\rm Diff}$, with a scalar product induced from the scalar product on ${\cal H}_{\rm kin}$. The space ${\cal H}^G_{\rm Diff}$ is then the physical Hilbert space of the quantum theory in both models.

The last ingredient to complete the quantization program is to define a quantum Hamiltonian operator which would generate the quantum dynamics through a Schr\"odinger-like equation
\be
i\hbar\frac{d}{dT}\ket\psi = \hat H\ket\psi
\ee
for any state $\ket\psi\in {\cal H}_{\rm Diff}^G$, where $T$ is the physical time equal to the value of the deparametrization field, either the scalar field or the dust field. This task can be achieved in a satisfactory manner through a careful regularization of the classical expressions of $\int_\Sigma d^3x\ h_{SF}(x)=:H_{SF}$ and $\int_\Sigma d^3x\ h_{D}(x)=:H_{D}$. Following \cite{AALM} for the massless scalar field case, and \cite{LQGSC} for the dust field case\footnote{In the present work we modify the Euclidean and Lorentzian parts of the Hamiltonian compared to the one introduced in \cite{LQGSC}: on one hand we change the ordering of the operators in the Lorentzian part, on the other other hand, instead of using Thiemann's trick in defining the Euclidean operator, we use the ``inverse volume'' operator in the final expression, in a similar way that is used in the curvature operator \cite{Curvature_op.}.}, in which different regularizations than the one due to Thiemann \cite{Thiemann96a, Thiemann98} were proposed, symmetric Hamiltonian operators acting on ${\cal H}^G_{\rm Diff}$ with a dense domain\footnote{Some for which there are proof of self-adjointness.} can be defined in both models. The ambiguity of the choice between various valid operators arises from the different available symmetric extensions of the non-symmetric operator derived from the regularization procedure. An ultimate criterion which could remove this ambiguity would be obtained through confronting the semi-classical physics induced by a certain choice of Hamiltonian operator with the predictions of the classical model.

In this article we will proceed with a specific choice of symmetric extension. Explicitly, the chosen Hamiltonian operators are as follows:
\begin{itemize}
 \item For the massless scalar field\footnote{Notice that the term $q^{ab}C^{gr}_aC^{gr}_b$ in \eqref{newscalar} is dropped from the physical Hamiltonian expression as it is assumed to vanish on spatial diffeomorphism invariant states, which is the case for states in the physical Hilbert space ${\cal H}^G_{\rm Diff}$. This reduces the second condition in \eqref{Cond1} on the gravitational constraints to $C^{gr}\leq0$, which can be implemented on the operator level by introducing an absolute value term as shown in the second line of \eqref{Sym.Hamil1}.}
\begin{align}\label{Sym.Hamil1}
\nonumber \hat{H}_{SF} &: \mathscr{D}(\hat{H}_{SF}) \subset {\cal H}^G_{\rm Diff}\ \longrightarrow \ {\cal H}^G_{\rm Diff}\\ 
 \hat{H}_{SF} &:=\sqrt{\frac{1+\beta^2}{16\pi G \beta^2}} \sum \limits_{x\in \Sigma} \sqrt{(\hat{C}_{x,SF}+\hat{C}_{x,SF}^{\dagger})\arrowvert_{\mathbbm{R}_{+}}}\\
 \nonumber&=\sqrt{\frac{1+\beta^2}{16\pi G \beta^2}} \sum \limits_{x\in \Sigma} \sqrt{\frac{1}{2}\left(\hat{C}_{x,SF}+\hat{C}_{x,SF}^{\dagger}+\abs{\hat{C}_{x,SF}+\hat{C}_{x,SF}^{\dagger}}\right)}\ ,
\end{align}
with
\begin{align}
\hat{C}_{x,SF}:= \frac{1}{1+\beta^2} \hat{C}_{x}^E + \hat{C}_{x}^L\ ,
\end{align}
such that
\begingroup
\allowdisplaybreaks
\noindent 
\begin{align}
\hat{C}_{x}^E :&=\kappa_1  \sum_{I,J} \epsilon_{ijk}\ \epsilon\left(\dot{e}_I,\dot{e}_J\right) \hat{h}^{k\ (l)}_{\alpha_{IJ}} \hat{J}_{x,e_I,i} \hat{J}_{x,e_J,j}\label{Euc}\ ,
\end{align}
\endgroup
\begingroup
\allowdisplaybreaks
\noindent 
\begin{align}
\hat{C}_{x}^L :&=\kappa_2 \sum \limits_{I,J}  \frac{\epsilon\left(\dot{e}_I,\dot{e}_J\right)}{2} \sqrt{\delta_{ii'} (\epsilon_{ijk} \hat{J}_{x,e_I,j} \hat{J}_{x,e_J,k}) (\epsilon_{i' j' k'} \hat{J}_{x,e_I,j'} \hat{J}_{v,x_J,k'})}\label{Lor} \\ \nonumber
&\qquad \qquad \times \left(\frac{2\pi}{\alpha} - \pi + \arccos \left[ \frac{ \delta_{kl} \hat{J}_{x,e_I,k} \hat{J}_{x,e_J,l}}{\sqrt{\delta_{kk'} \hat{J}_{x,e_I,k} \hat{J}_{x,e_I,k'}} \sqrt{\delta_{ll'} \hat{J}_{x,e_J,l} \hat{J}_{x,e_J,l'}}} \right]\right)\ ,
\end{align}
\endgroup
where $\mathscr{D}(\hat{H}_{SF})$ is the domain of the operator $\hat{H}_{SF}$ which contains the span of the spin network basis. The sum in \eqref{Sym.Hamil1} is over the all points $x$ of $\Sigma$, but it reduces to a finite sum over the vertices of a graph when the operator acts on a spin network state. $\hat{C}_{x,SF}^{\dagger}$ is the adjoint operator of $\hat{C}_{x,SF}$. The operators $\hat{C}_{x}^E$ and $\hat{C}_{x}^L$ represent the Euclidean and Lorentzian parts of the Hamiltonian operator respectively. The operator $\hat{C}_{x}^L$ is graph-preserving while the operator $\hat{C}_{x}^E$ is graph-changing. Here $\kappa_1$ and $\kappa_2$ are averaging coefficient and $\alpha$ an unfixed constant, resulting from the regularization procedures \cite{LQGSC, AALM, Curvature_op.}. The two sums in \eqref{Euc} and \eqref{Lor} are over pairs of curves $\{e_I,e_J\}$ meeting at a point $x$ with tangent vectors $\{\dot e_I,\dot e_J\}$. The coefficients $\epsilon\left(\dot{e}_I,\dot{e}_J\right)$ is $0$ if $\dot{e}'_I$ and $\dot{e}'_J$ are linearly dependent or $1$ otherwise. It is important to note that the square root present in the definition of the operator $\hat{H}_{SF}$ is to be understood as taking the square root of the operator $\hat{C}_{x,SF}+\hat{C}_{x,SF}^{\dagger}$ restricted to the positive part of its spectrum.

\item For the dust field
\begin{align}\label{Sym.Hamil2}
\nonumber \hat{H}_{D} &: \mathscr{D}(\hat{H}_{D}) \subset {\cal H}^G_{\rm Diff}\ \longrightarrow \ {\cal H}^G_{\rm Diff}\\ 
 \hat{H}_{D} &:=\frac{1+\beta^2}{32\pi G \beta^2} \sum \limits_{x\in \Sigma} \hat{C}_{x,D}+\hat{C}_{x,D}^\dagger
\end{align}
with
\begin{align}
\hat{C}_{x,D}:=  \frac{1}{1+\beta^2} \sqrt{\widehat{V^{-1}}}\hat{C}_{x}^E\sqrt{\widehat{V^{-1}}}+\sqrt{\widehat{V^{-1}}}\hat{C}_{x}^L\sqrt{\widehat{V^{-1}}}\ ,
\end{align}
where $\widehat{V^{-1}}$ is the ``inverse volume'' operator \cite{lunghezza2} defined in terms of the LQG volume operator \cite{AshtekarLewand98} as\footnote{In other words, given the spectral decomposition of the volume operator  $\hat{V}=\sum_i v_i \ket{v_i}\bra{v_i}$, we have
\begin{align}
\qquad \widehat{V^{-1}}\ket{v_i}=\left\{ \begin{array}{cl}
 v_i^{-1} \ket{v_i} & \text{if $v_i \neq 0$}, \\ 0 &\text{otherwise.} \end{array} \right.
\end{align}}
\begin{align}
 \widehat{V^{-1}}:= \lim \limits_{s \rightarrow 0} \bigl(\hat{V}^2 + s^2 l_p^6 \bigr)^{-1} \hat{V}.
\end{align}
$\mathscr{D}(\hat{H}_{D})$ is the domain of the operator $\hat{H}_{D}$ which contains the span of the spin network basis and $\hat{C}_{x,D}^{\dagger}$ is the adjoint operator of $\hat{C}_{x,D}$.
\end{itemize}
Schematically, given a spin network state with a closed graph $\Gamma$, the two operators $\hat{C}_{x}^E$ and $\hat{C}_{x}^L$ defined in \eqref{Euc} and \eqref{Lor} act on the vertex $x$ of the graph $\Gamma$ as follows:
\begingroup
\allowdisplaybreaks
\begin{align}
 \hat{C}_{x}^E \quad \makeSymbol{\includegraphics[scale=1.75]{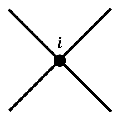}} \quad &= \quad \makeSymbol{\includegraphics[scale=1.75]{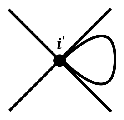}} \quad + \quad \makeSymbol{\includegraphics[scale=1.75]{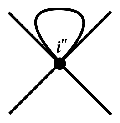}} \quad +\quad \dots\qquad ,\\
 \hat{C}_{x}^L \quad \makeSymbol{\includegraphics[scale=1.75]{node.eps}} \quad &= \quad \makeSymbol{\includegraphics[scale=1.75]{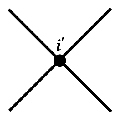}}\qquad .
\end{align}
\endgroup

\section{Approximation methods for LQG dynamics}\label{sec:approximation}
Having fully completed the quantization of both systems, we naturally turn to the question of testing the quantum theories. An important aspect of this question is to obtain a sufficient control on the quantum dynamics. More specifically, we need to be able to compute transition amplitudes and the evolution of observables with a relatively good precision. Considering the Hamiltonian operators on the non-separable physical Hilbert space ${\cal H}^G_{\rm Diff}$ defined above, a derivation of their complete spectral decomposition has not been achieved so far. It becomes then imperative to develop and use approximate methods in the analysis of the dynamics. 

\subsection{Expansion of expectation values in powers of time}

As mentioned in the introduction, one can consider the expansion in powers of the time variable of the evolution operator and introduce a truncation of the expansion at a certain fixed order of time. Such truncation forms a valid approximation of the time evolution when the time interval under consideration is sufficiently small. We investigate this approach through a couple of examples within the dust field model in section \ref{sec32}. In this context, the most convenient way to compute the time evolution of the expectation value of an operator $A$ is by evaluating the coefficients in the power series expansion of the expectation value,
\be\label{ev-1}
\langle A(T)\rangle = \sum_n a_n T^n \ .
\ee
The coefficients are given by expectation values of repeated commutators of $A$ with the Hamiltonian in the initial state $\ket{\psi_0}$,
\be\label{ev-2}
a_n = \frac{(-i)^n}{n!}\bigl\langle \underbrace{[H,\dots,[H,[H,A]]\dots]}_{n\;\text{commutators}}\bigr\rangle_{\psi_0}\ .
\ee
The advantage of directly considering the expansion of an expectation value $\langle A(T)\rangle$, as opposed to computing the evolved state vector $\ket{\Psi(T)}$ truncated at some order $T^{n}$, is that the expectation value can be determined up to order $T^{n}$ without having to compute all the components of the truncated state vector $\ket{\Psi(T)}$ in the spin-network basis. For example, if the initial state is based on a single graph containing no ``special loops'' (of the kind created by the Euclidean part of the Hamiltonian), and the operator $A$ is graph-preserving\footnote{Which is the case in the examples we present in Section \ref{sec32}.}, and one wants to find the expectation value $\langle A(T)\rangle$ to order $T^n$, then states containing more than $\lfloor n/2 \rfloor$ special loops do not enter the calculation of the numbers $a_k\ (k=1,\dots,n)$, even though the state vector $\ket{\Psi(T)}$ truncated at order $n$ has components containing up to $n$ special loops.


However, this method is not appropriate to deal with the Hamiltonian operator $\hat{H}_{SF}$ present in the model with a massless scalar field \eqref{Sym.Hamil1}. The reason being the presence of the square root in the expression of $\hat{H}_{SF}$, which requires an access to the spectral decomposition of the operator under the square root.

\subsection{Perturbation theory with the Barbero-Immirzi parameter}

An alternative solution to the problem is provided by standard time-independent perturbation theory of quantum mechanics. In the following we introduce a perturbative expansion of the Hamiltonian, the small perturbation parameter being determined by the Barbero-Immirzi parameter $\beta$. This approach allows us to define an approximate spectral decomposition of the physical Hamiltonian $\hat{H}_{SF}$, and hence the time-evolution operator $U_{SF}(T)$, on appropriate time intervals.\\

Recall that the expression of the operator $\hat{C}_{x,SF}$ is
\begin{align}
\hat{C}_{x,SF}:= \frac{1}{1+\beta^2} \hat{C}_{x}^E + \hat{C}_{x}^L\ ,
\end{align}
and hence\footnote{The operator $\hat{C}_{x}^L$ and the curvature operator $\sqrt{\widehat{V^{-1}}}\hat{C}_{x}^L\sqrt{\widehat{V^{-1}}}$ are self-adjoint operators \cite{Curvature_op.}, therefore
\begin{align}
 \hat{C}_{x}^{L\ \dagger}=\hat{C}_{x}^L\ ,\qquad (\sqrt{\widehat{V^{-1}}}\hat{C}_{x}^L\sqrt{\widehat{V^{-1}}})^\dagger=\sqrt{\widehat{V^{-1}}}\hat{C}_{x}^L\sqrt{\widehat{V^{-1}}}
\end{align}}
\begin{align}
 \hat{C}_{x,SF}+\hat{C}_{x,SF}^{\dagger} :=  \frac{1}{1+\beta^2} (\hat{C}_{x}^E+\hat{C}_{x}^{E\ \dagger}) + 2 \hat{C}_{x}^L\ .
\end{align}
Since the operator $\hat{C}_{x}^L$ is graph preserving and acts locally on the vertices of the graph without changing the $SU(2)$ representations \cite{Curvature_op.,AALM}, its spectral decomposition breaks down to stable finite dimensional blocks. Each block corresponds to the Hilbert space of a fixed graph with fixed coloring (spins) and takes the form of a tensor product over the vertices of stable sub-blocks, each representing a separate intertwiner space assigned to each vertex of the colored graph. Given a colored graph, the dimension of each intertwiner space is then fixed, hence one can proceed with the diagonalization of the (self-adjoint) Lorentzian part of the Hamiltonian that is the operator $\hat{C}_{x}^L$. 

Having the spectral decomposition of this operator, the idea is to treat the Euclidean part of the operator, $\frac{1}{1+\beta^2} (\hat{C}_{x}^E+\hat{C}_{x}^{E\ \dagger})$, as a perturbation to the Lorentzian part with $1/(1+\beta^2)$ being the perturbation parameter. This means that we will assume that the Barbero-Immirzi parameter is significantly large, $\beta\gg 1$, large enough so that the perturbative expansion in $1/(1+\beta^2)$ gives a good approximation for the eigenvalues and eigenstates\footnote{Since we expect to be dealing with unbounded operators, it is not clear to us yet if, given a fixed value of $\beta$, the perturbative expansion would be valid for all eigenstates of $\hat{C}_{SF,\text{sym},x}^L$ or $\hat{C}_{D,\text{sym},x}^L$ on ${\mathscr H}^G_\text{Diff}$.
} of the Hamiltonian. The condition for this is that the corrections to the eigenvalues and eigenvectors should be small in norm, compared to the corresponding eigenvalues and eigenvectors of the unperturbed Hamiltonian (in our case, the Lorentzian part of the Hamiltonian). 

The procedure is then as follows: given an intertwiner space ${\cal I}_v$ of dimension $d_v$ associated to a vertex $v$ of a given colored graph, the Lorentzian part operator is put in a diagonal form
\begin{align}
 2\hat{C}_{v}^L=\sum \limits_{i=1}^{d_v} \lambda_i \ket{\lambda_i}\bra{\lambda_i}\ .
\end{align}
For $\beta$ sufficiently large, we can write
\begin{align}
2\hat{C}_{v}^L +\frac{1}{1+\beta^2} (\hat{C}_{v}^E+\hat{C}_{v}^{E\ \dagger})= \sum \limits_{i=1}^{d_v} \lambda_i^\prime \ket{\lambda_i^\prime}\bra{\lambda_i^\prime}\ ,
\end{align}
and replace the eigenvalues $\lambda_i'$ and the eigenstates $\ket{\lambda_i'}$ with their approximate expressions given by perturbation theory to second order in $1/(1+\beta^2)$. We have
\begin{align}
 \lambda_i^\prime =\lambda_i + \biggl(\frac{1}{1+\beta^2}\biggr)^2 \sum \limits_{\substack{k=1 \\ \lambda_k\neq \lambda_i}}^{d'_v} \frac{\bigl|\bra{\lambda_i}\hat{C}_{v}^E+\hat{C}_{v}^{E\ \dagger}\ket{\lambda_k}\bigr|^2}{\lambda_i-\lambda_k}+{\cal O}\left((1+\beta^2)^{-3}\right)\ ,\label{Eig1}
\end{align}
and
\begin{align}
  \ket{\lambda_i^\prime} =\ket{\lambda_i} &+ \frac{1}{1+\beta^2} \sum \limits_{\substack{k=1 \\ \lambda_k\neq \lambda_i}}^{d'_v} 
	\frac{\bra{\lambda_k}\hat{C}_{v}^E+\hat{C}_{v}^{E\ \dagger}\ket{\lambda_i}}{\lambda_i-\lambda_k}\ket{\lambda_k}  \notag \\ 
	&+\biggl(\frac{1}{1+\beta^2}\biggr)^2 \sum \limits_{\substack{k=1 \\ \lambda_k\neq \lambda_i}}^{d'_v} \biggl(\sum \limits_{\substack{n=1 \\ \lambda_n\neq \lambda_i}}^{d'_v} \frac{\bra{\lambda_k}\hat{C}_{v}^E+\hat{C}_{v}^{E\ \dagger}\ket{\lambda_n}\bra{\lambda_n}\hat{C}_{v}^E+\hat{C}_{v}^{E\ \dagger}\ket{\lambda_i}}{(\lambda_i-\lambda_k)(\lambda_i-\lambda_n)}\biggr) \ket{\lambda_k} \notag \\
	&+\biggl(\frac{1}{1+\beta^2}\biggr)^2\biggl(-\frac{1}{2}\sum \limits_{\substack{k=1 \\ \lambda_k\neq \lambda_i}}^{d'_v} \frac{\bigl|\bra{\lambda_k}\hat{C}_{v}^E+\hat{C}_{v}^{E\ \dagger}\ket{\lambda_i}\bigr|^2}{(\lambda_k-\lambda_i)^2}\biggr)\ket{\lambda_i}  + {\cal O}\left((1+\beta^2)^{-3}\right)\label{Eig2} \ .
\end{align}
Because the Euclidean part does not preserve each of the stable subspaces of the Lorentzian part separately, as it modifies the graph structure at the vertex $v$, the first-order correction to the eigenvalue $\lambda_i$ vanishes. Also, the sums in \eqref{Eig1} and \eqref{Eig2} are over the eigenstates of the Lorentzian part in the new intertwiner spaces at $v$, which together contain the image of the space ${\cal I}_v$ by the Euclidean part. The upper limit of the summation $d'_v$ is then the finite sum of dimensions of the new intertwiner spaces at the vertex $v$.

The derivation of the corrections to the eigenstate requires some care, due to a degeneracy of the unperturbed operator that is not removed by the perturbation (at least to second order in the perturbation parameter), and is therefore discussed in the Appendix.

It is then straightforward to obtain the explicit expression of the square root operator and the evolution operator:
\begin{align}
\sqrt{(\hat{C}_{v,SF}+\hat{C}_{v,SF}^{\dagger})\arrowvert_{\mathbbm{R}_{+}}}&=\sqrt{\frac{1}{2}\left(\hat{C}_{v,SF}+\hat{C}_{v,SF}^{\dagger}+\abs{\hat{C}_{v,SF}+\hat{C}_{v,SF}^{\dagger}}\right)}=\sum \limits_{\substack{i=1 \\ \lambda_i^\prime \geq 0}}^{d_v} \sqrt{ \lambda_i^\prime} \ket{\lambda_i^\prime}\bra{\lambda_i^\prime}\ ,\\
U_{SF}(T) := \exp\biggl(- \frac{i}{\hbar} T \hat{H}_{SF}\biggr) &=  \prod \limits_{x\in \Sigma} \exp\biggl(- \frac{i}{\hbar} T \sqrt{\frac{1+\beta^2}{16\pi G \beta^2}}\sqrt{\hat{C}_{x,SF}+\hat{C}_{x,SF}^{\dagger}}\biggr) \notag \\
&= \prod \limits_{x\in \Sigma}\ \sum \limits_{\substack{i=1 \\ \lambda_i^\prime \geq 0}}^{d_x} \exp\biggl(- \frac{i}{\hbar} T \sqrt{\frac{(1+\beta^2)}{16\pi G \beta^2}\lambda_i^\prime}\biggr)\ \ket{\lambda_i^\prime}\bra{\lambda_i^\prime}\ .\label{Evol.Op}
\end{align}
It follows that given an operator $A$ and an initial state $\ket{\Psi_0}$, the state at time $T$ is given by $\ket{\Psi(T)} = U_{SF}(T)\ket{\Psi_0}$ and the expectation value $\mean{A(T)}$ is computed as
\begin{align}\label{Gen.Exp.}
 \mean{A(T)}&=\bra{\Psi(T)}A\ket{\Psi(T)}\\ \nonumber
 &=\prod \limits_{x\in \Sigma}\ \sum \limits_{\substack{i,j=1 \\ \lambda_i^\prime \geq 0 \\ \lambda_j^\prime \geq 0}}^{d_x} \exp\biggl(- \frac{i}{\hbar} T \sqrt{\frac{(1+\beta^2)}{16\pi G \beta^2}}\left(\sqrt{\lambda_i^\prime}-\sqrt{\lambda_j^\prime}\right)\biggr)\ \langle\Psi_0|\lambda_j^\prime\rangle \bra{\lambda_j^\prime}A \ket{\lambda_i^\prime}\langle\lambda_i^\prime|\Psi_0\rangle\ .
\end{align}
In order to compute the expectation value of the volume or the curvature operator to second order in $1/(1+\beta^2)$, some parts of the expression \eqref{Eig2} for the corrected state vector can be discarded, because they do not contribute to the expectation value at the specified order in the perturbation. If the initial state $\ket{\Psi_0}$ is based on a single graph $\Gamma_0$, which will be the case in the examples in the following section, then the following simplifications can be made. For unperturbed eigenstates based on $\Gamma_0$, one has to take the first-order correction, and the part of the second-order correction which is based on $\Gamma_0$. For unperturbed eigenstates whose graph is $\Gamma_0$ decorated with one special loop, it suffices to take the part of the first-order correction based on $\Gamma_0$, and the second-order correction can be discarded entirely. Unperturbed eigenstates whose graph contains more than one special loop do not enter the calculation at second order in $1/(1+\beta^2)$. A more detailed expression of the expectation value \eqref{Gen.Exp.}, up to the second order in perturbation theory, is given in Appendix \ref{app2}.

All that was mentioned above for the operator $\hat{H}_{SF}$ can be similarly applied to the operator $\hat{H}_{D}$ in the dust model. Later in the examples within the dust model, we separately test the Barbero-Immirzi parameter pertubative expansion (the $\beta$-expansion), and the approximation obtained by the short time truncation in the time expansion of the evolution operator.

\clearpage

\section{Examples and numerical analysis}\label{sec:examples}

In the following graphics we present the evolution of the expectation values of the volume operator and the curvature operator \cite{Curvature_op.}. In the scalar field deparametrized model we use the $\beta$-expansion with certain values of $\beta$, while in the dust model we consider both the $\beta$-expansion and the time expansion approximation. We consider initial states corresponding to certain eigenvectors of the volume operator with a graph consisting of a single non-degenerate $4$-valent vertex $v$.

In all the calculations, we fix all the constants in the operators as follows
\begin{align}
 16\pi G=\hbar=\kappa_0=1\ ,\ \alpha=3\ ,
\end{align}
where $\kappa_0$ is the averaging constant present in the definition of the volume operator \cite{AshtekarLewand98}. Additionally, the $SU(2)$ representation of the holonomies associated to the special loops created by the Euclidean part operator is fixed to $1/2$.

\subsection{Perturbation theory in the scalar field and dust field models}

The $\beta$-expansion is taken to second order, because with our choice of initial states, the Euclidean part of the Hamiltonian does not contribute to the time evolution of the expectation values of volume and curvature at first order of the expansion. The expectation values $\langle V(T)\rangle$ and $\langle R(T)\rangle$ are computed from equation \eqref{Gen.Exp.}, with the eigenvalues $\lambda_i'$ and eigenstates $\ket{\lambda_i'}$ being given by equations \eqref{Eig1} and \eqref{Eig2}.

In all the graphics below, the parameter $T$ stands for the standard time given either by the scalar field or the dust field depending on the considered case. The parameters $T'$ and $T''$ in the embedded graphics stand for the rescaled times given by
\begin{align}
 T':=\sqrt{1+\beta^2}\ T\qquad ,\qquad T'':=\frac{1+\beta^2}{\abs{\beta}^{3/2}}\ T\ .
\end{align}


\begin{itemize}
 \item Perturbation theory in the scalar field model:
 \begin{itemize}
 \item Eigenvectors with spins $j=2$:
 
 \begin{figure}[!h]
  \centering
  \includegraphics[width=0.6\textwidth]{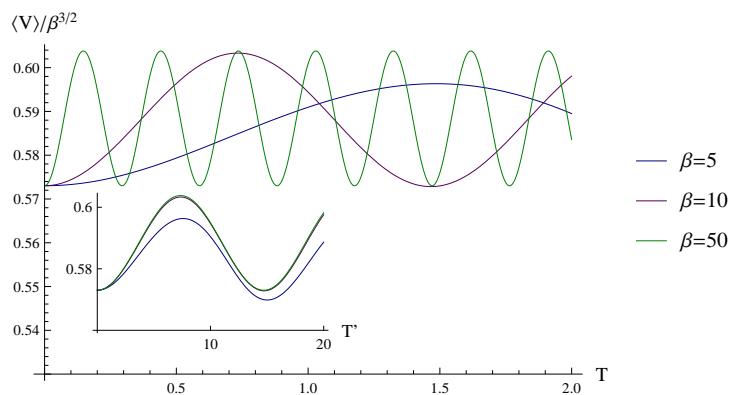}
  \caption{Evolution of the expectation value $\mean{V}$ of the volume operator with an initial eigenvector with eigenvalue $v=0.5730$.}
  \label{VS2B}
 \end{figure}
 
 \clearpage
 
 \begin{figure}[!h]
  \centering
  \includegraphics[width=0.6\textwidth]{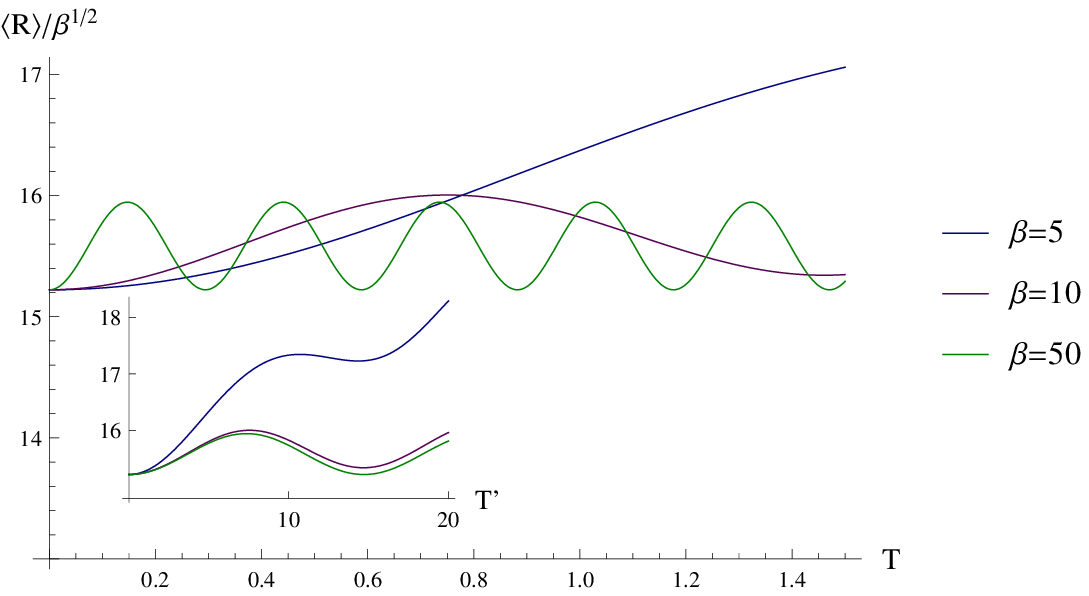}
  \caption{Evolution of the expectation value $\mean{R}$ of the curvature operator with an initial eigenvector with eigenvalue $v=0.8725$.}
  \label{RS2B}
 \end{figure}
 
 
 \item Eigenvectors with spins $j=10$:
 
  \begin{figure}[!h]
  \centering
  \includegraphics[width=0.6\textwidth]{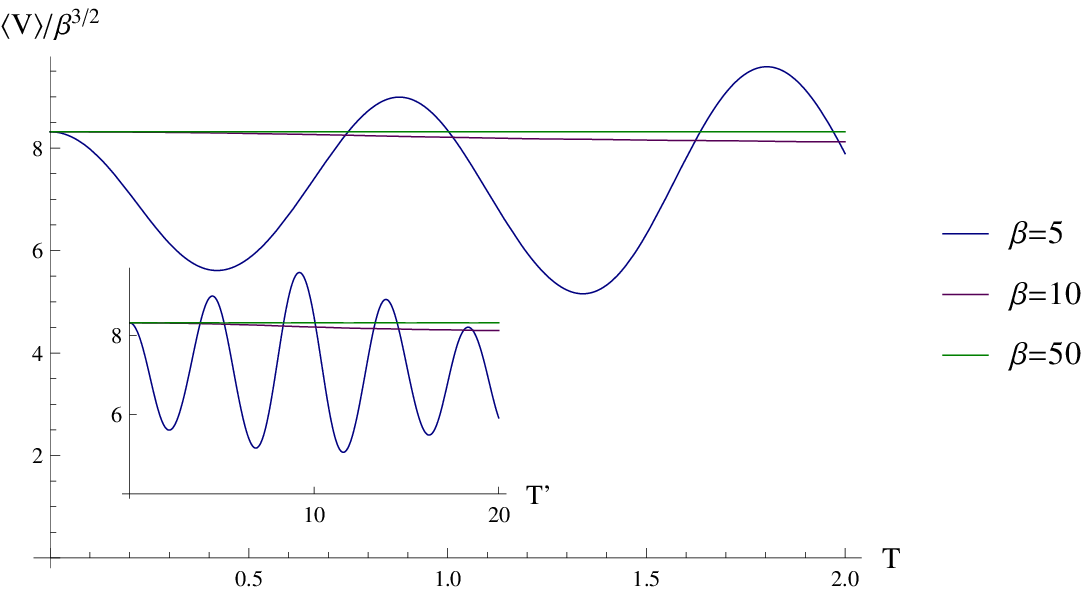}
  \caption{Evolution of the expectation value $\mean{V}$ of the volume operator with an initial eigenvector with eigenvalue $v=8.3177$.}
  \label{V1S10B}
 \end{figure}
 
 
 \begin{figure}[!h]
  \centering
  \includegraphics[width=0.6\textwidth]{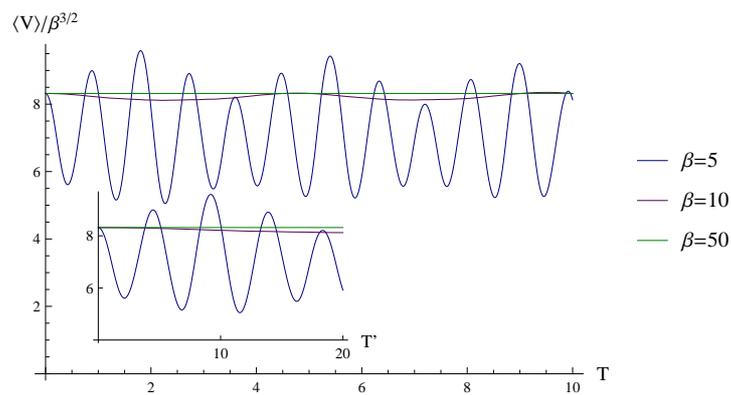}
  \caption{Evolution of the expectation value $\mean{R}$ of the curvature operator with an initial eigenvector with eigenvalue $v=5.1078$.}
  \label{R1S10B}
 \end{figure}

 \end{itemize}
 
  \clearpage
 
 \item Perturbation theory in the dust field model:
 \begin{itemize}
 \item Eigenvectors with spins $j=2$:
 
 \begin{figure}[!h]
  \centering
  \includegraphics[width=0.6\textwidth]{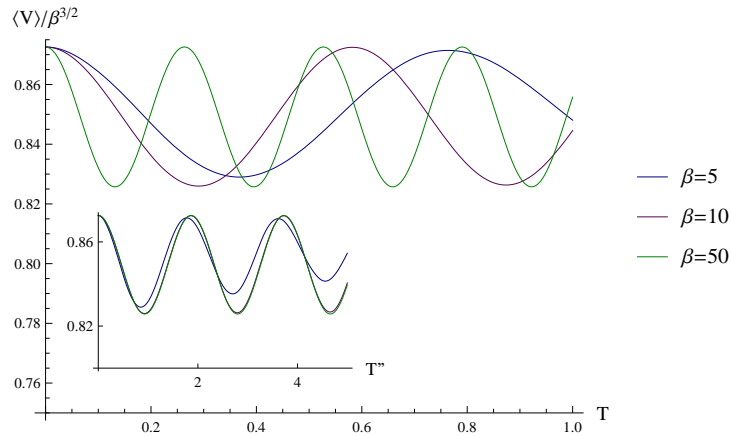}
  \caption{Evolution of the expectation value $\mean{V}$ of the volume operator with an initial eigenvector with eigenvalue $v=0.8725$.}
  \label{VD2B}
 \end{figure} 
 
 
  \begin{figure}[!h]
  \centering
  \includegraphics[width=0.6\textwidth]{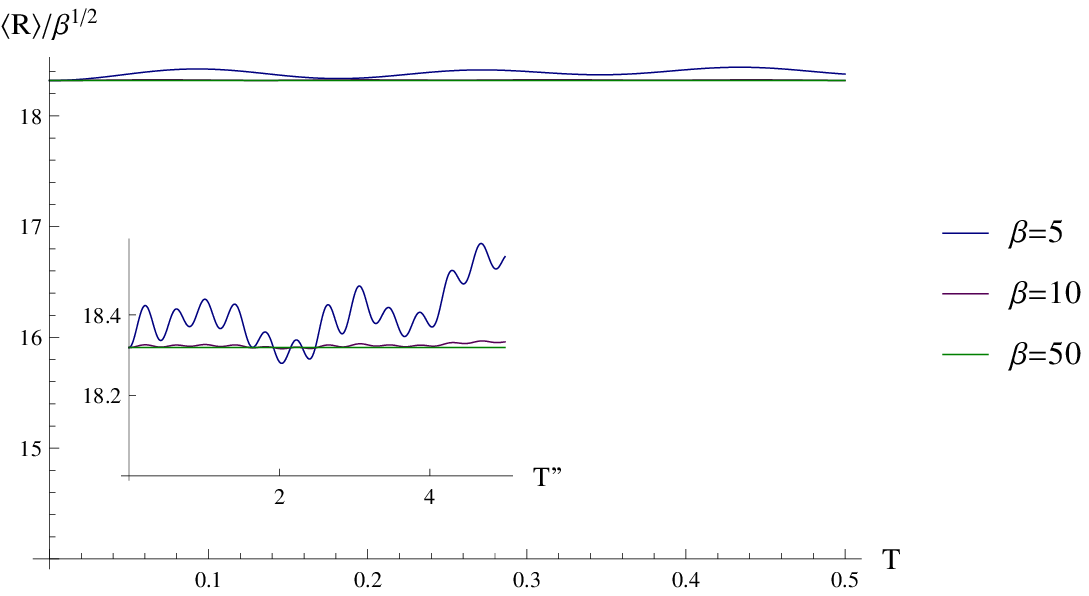}
  \caption{Evolution of the expectation value $\mean{R}$ of the curvature operator with an initial eigenvector with eigenvalue $v=0.5730$.}
  \label{RD2B}
 \end{figure}
 

 \item Eigenvectors with spins $j=10$:
 
 \begin{figure}[!h]
  \centering
  \includegraphics[width=0.6\textwidth]{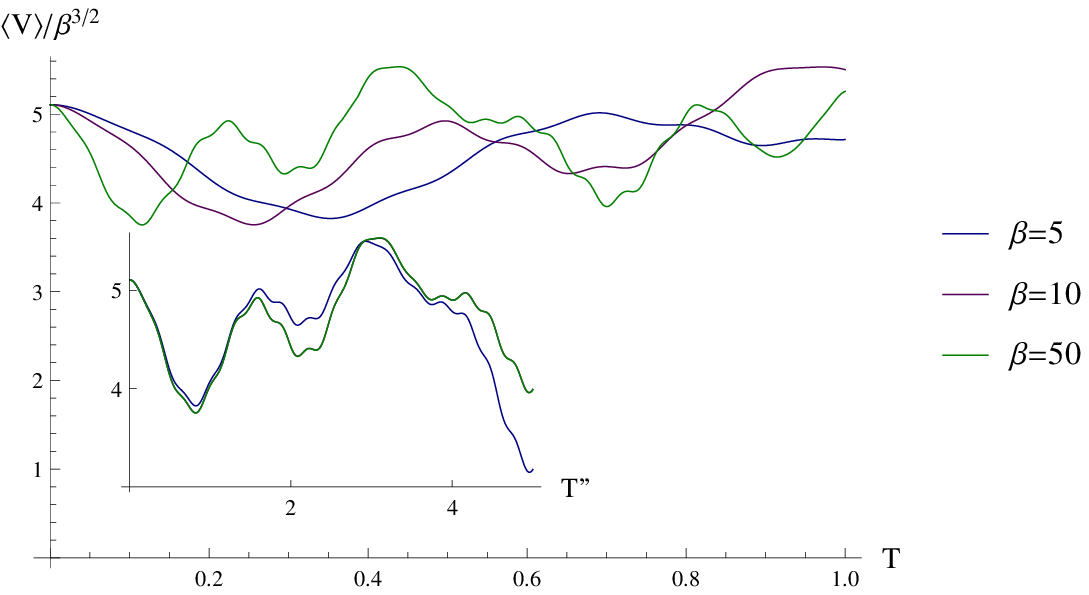}
  \caption{Evolution of the expectation value $\mean{V}$ of the volume operator with an initial eigenvector with eigenvalue $v=5.1078$.}
  \label{V1D10B}
 \end{figure}
 
 \clearpage

 \begin{figure}[!h]
  \centering
  \includegraphics[width=0.6\textwidth]{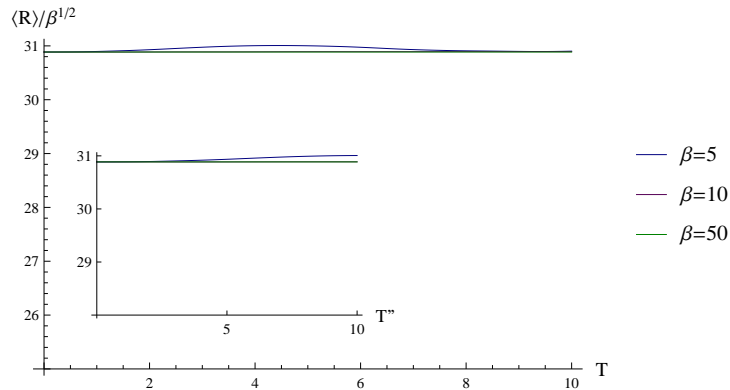}
  \caption{Evolution of the expectation value $\mean{R}$ of the curvature operator with an initial eigenvector with eigenvalue $v=8.3177$.}
  \label{R1D10B}
 \end{figure}
 
 \end{itemize}
\end{itemize}

Discussion:
\begin{itemize}

 \item Since the initial states we are considering in this numerical analysis correspond to specific spin-network states, and the evolution operators in this approximation contain a finite order of the graph changing Euclidean operators, the expectation values of the volume and curvature operators are both bounded throughout the time evolution of those states. It is possible that unbounded expectation values can be obtained in this approximation by considering initial states which take the form of infinite linear combination of spin-network states.

 \item The degeneracy of the volume and curvature eigenvalues is preserved under time evolution, in the sense that two degenerate initial states give rise to the same function $\mean{V(T)}$ or $\mean{R(t)}$. Furthermore, the degeneracy present in the eigenvalues of the Lorentzian part of the Hamiltonian is not removed by the perturbation provided by the Euclidean part, at least to second order in perturbation theory, suggesting that the degeneracy might be preserved exactly. These observations strongly indicate the existence of some symmetry shared by the volume operator and the Lorentzian and Euclidean operators. 

 \item The large fluctuations of the expectation values curves for $\beta=5$ with respect to $\beta=50$ (which can be seen as the limit where the perturbations is totally negligible) in the figures above demonstrate that the value $\beta=5$ of the Barbero-Immirzi parameter is not good enough to make sense of the perturbation method, suggesting that, at least according to these examples, the range of $\beta$ consistent with the pertubative treatment is $\abs{\beta} \gtrsim 10$.

 \item In the case of the scalar field model, when the perturbation from the Euclidean part is small (e.g. spin $2$ case, or \ref{R1S10B} for spin $10$), one can notice a periodic evolution of the expectation values of the volume and curvature operators. This periodicity seems to manifest for all eigenvectors of the volume operator, independently of the intertwiner space. This is another piece of evidence pointing towards the presence of a certain symmetry between the volume operator, the curvature operator and the Lorentzian operator $\hat C_{SF}$. We leave the investigation of the symmetry properties of our operators as a question for future study.
 
 \item Figures \ref{RD2B} and \ref{R1D10B}, for the expectation value of the curvature operator in the dust model, show practically constant curves for $\beta=10$ and $\beta=50$. This is expected because when the perturbation is small, the dust model Hamiltonian reduces to almost the curvature operator itself, hence the constant expectation value.
 
 \item Finally, the embedded graphics on the right of each figure display the evolution with respect to the rescaled time. Comparing those graphics to the graphics for the evolution with respect to the standard time exhibits how the overall factors depending on $\beta$ in the evolution operators affect the phases in the evolution curves. Those overall factors are obtained by factorizing out all the dependence on $\beta$ in the Lorentzian and Euclidean parts of the Hamiltonian operator, i.e one write the Hamiltonian in the form
 \begin{align}
  \hat H= f(\beta) \left(\hat{C}_0^L + \frac{1}{1+\beta^2} \hat{C}_0^E\right)\ ,
 \end{align}
 such that $\hat{C}_0^L$ and $\hat{C}_0^E$ are independent of $\beta$. $f(\beta)$ is then the rescaling factor which equals $\sqrt{1+\beta^2}$ for $\hat H_{SF}$ and $(1+\beta^2)/\abs{\beta}^{3/2}$ for $\hat H_{D}$. 
 
\end{itemize}


\subsection{Time-expansion approximation in the dust field model}\label{sec32}

The time-expansion in the following examples is taken up to fourth order. The expectation values of the volume and curvature operators are computed according to eqs. \eqref{ev-1} and \eqref{ev-2}. At order $T^4$, the set of graphs that enters the computation consists of the graph of the initial state (a single four-valent node), and of the graphs generated by no more than two actions of the Euclidean part of the Hamiltonian on the initial state.

The computation of the coefficients of the power series expansion of the expectation values $\langle V(T)\rangle$ and $\langle R(T)\rangle$ reveals the following properties:
\begin{itemize}
\item Only even powers of $T$ are present in the expansion of the functions $\langle V(T)\rangle$ and $\langle R(T)\rangle$. The coefficients of the odd powers of $T$ ($T^1$ and $T^3$) vanish up to numerical rounding error. This seems to suggest the invariance of the Hamiltonian, the volume and curvature operators under time reversal.
\item Degeneracy of eigenvalues is again preserved under time evolution, in the sense that for a given degenerate eigenvalue of the volume or the curvature, the function $\langle V(T)\rangle$ or $\langle R(T)\rangle$ does not depend on which eigenstate belonging to the degenerate eigenvalue is selected as the initial state.
\end{itemize}

To determine the range of validity of the time expansion, one should estimate the value of $T$ at which the magnitude of the first neglected term in the expansion of an expectation value (we expect this to be the term of order $T^6$) starts being comparable to the terms included in the approximation. This criterion can be tested in a toy example in which the Hamiltonian consists only of the Lorentzian part, and the dynamics can be evaluated exactly. In this case we find that the criterion correctly predicts the order of magnitude of the time at which an expectation value computed from the fourth-order time expansion begins to diverge from the exact expectation value.
 \clearpage
\begin{itemize}
 \item Eigenvectors with spins $j=2$:
 
 \begin{figure}[!h]
  \centering
  \includegraphics[width=0.6\textwidth]{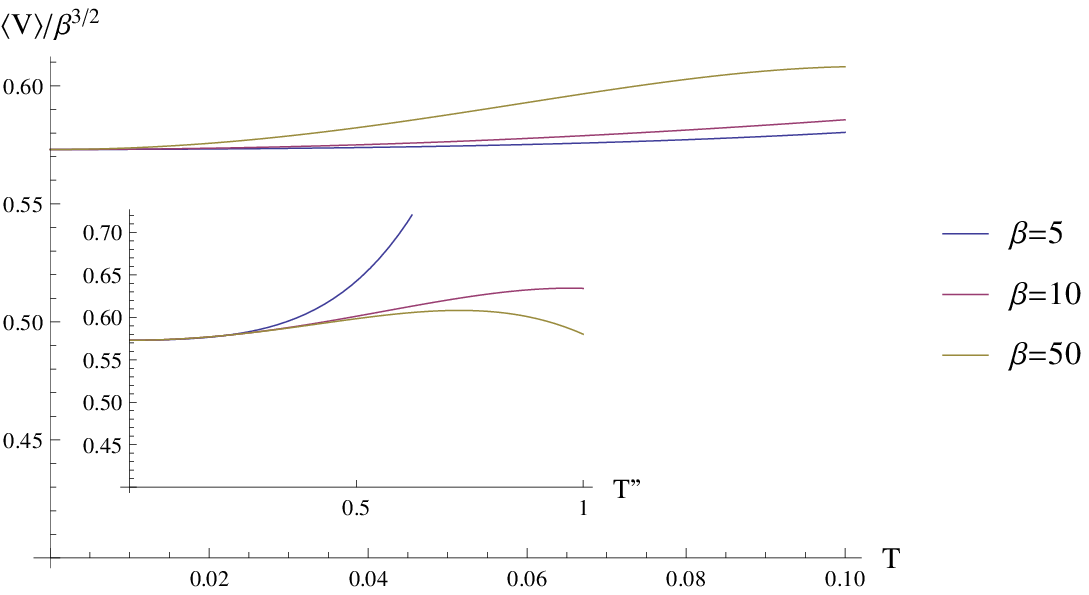}
  \caption{Evolution of the expectation value $\mean{V}$ of the volume operator with an initial eigenvector with eigenvalue $v=0.5730$.}
  \label{EVS2B}
 \end{figure} 
 
 
 \begin{figure}[!h]
  \centering
  \includegraphics[width=0.6\textwidth]{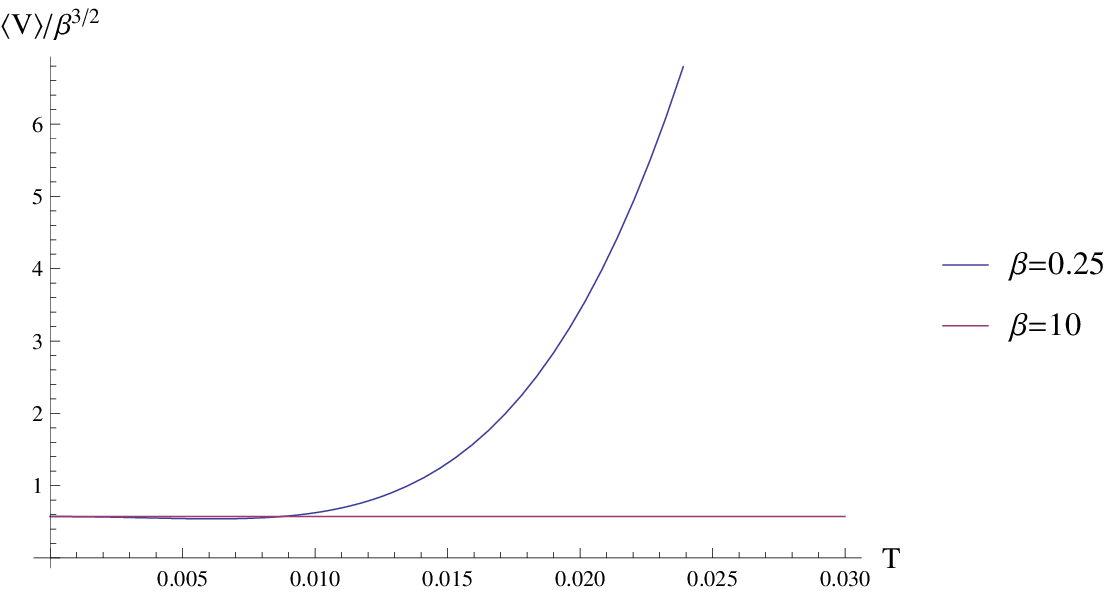}
  \caption{Evolution of the expectation value $\mean{V}$ of the volume operator with an initial eigenvector with eigenvalue $v=0.5730$.}
  \label{EVS2B0}
 \end{figure}


 \begin{figure}[!h]
  \centering
  \includegraphics[width=0.6\textwidth]{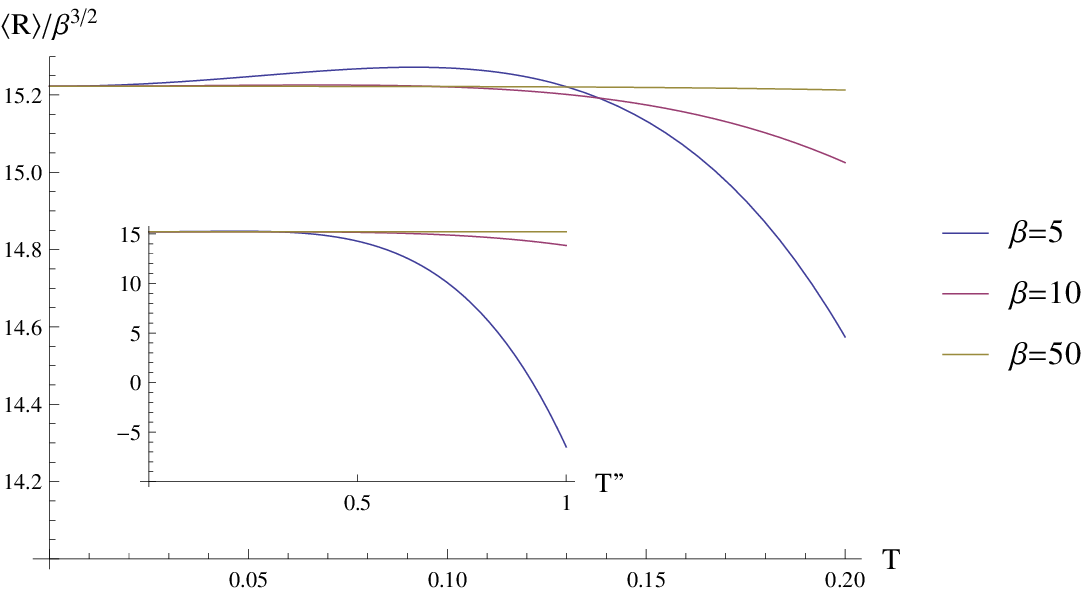}
  \caption{Evolution of the expectation value $\mean{R}$ of the curvature operator with an initial eigenvector with eigenvalue $v=0.8725$.}
  \label{ERS2B}
 \end{figure}

 \clearpage
 
 \begin{figure}[!h]
  \centering
  \includegraphics[width=0.6\textwidth]{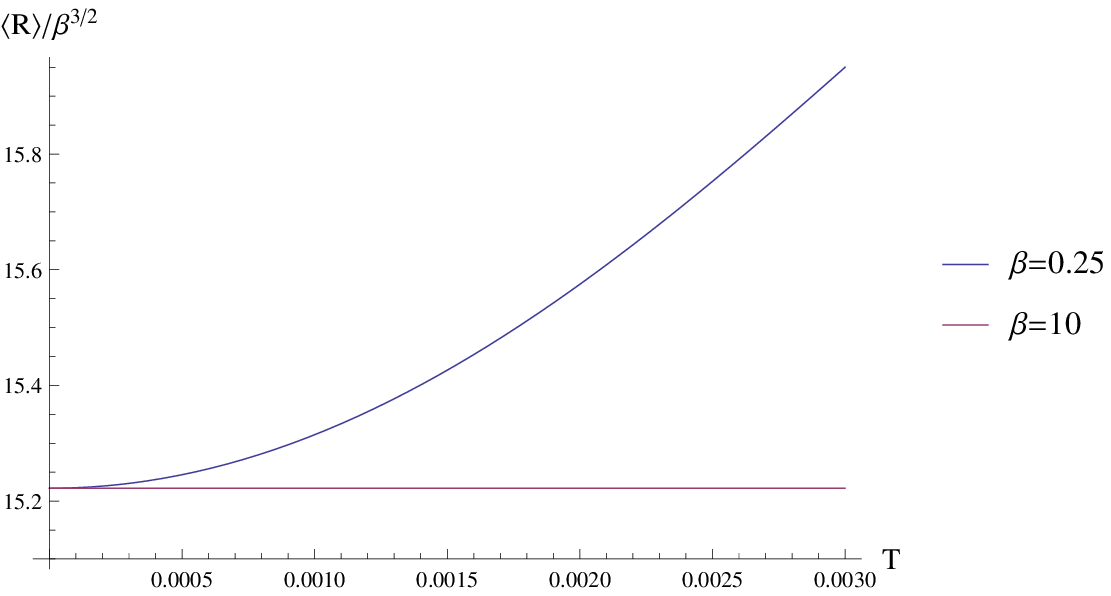}
  \caption{Evolution of the expectation value $\mean{R}$ of the curvature operator with an initial eigenvector with eigenvalue $v=0.8725$.}
  \label{ERS2B0}
 \end{figure}
 
 
  \item Eigenvectors with spins $j=25$:

 \begin{figure}[!h]
  \centering
  \includegraphics[width=0.6\textwidth]{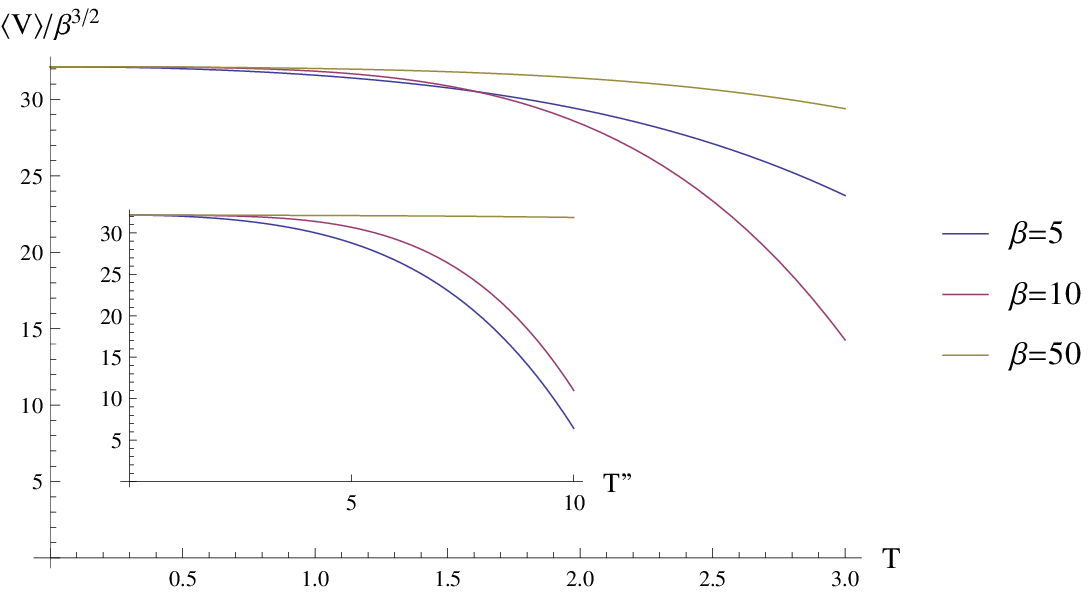}
  \caption{Evolution of the expectation value $\mean{V}$ of the volume operator with an initial eigenvector with eigenvalue $v=32.1396$.}
  \label{EVS25B}
 \end{figure}
 
 
 \begin{figure}[!ht]
  \centering
  \includegraphics[width=0.6\textwidth]{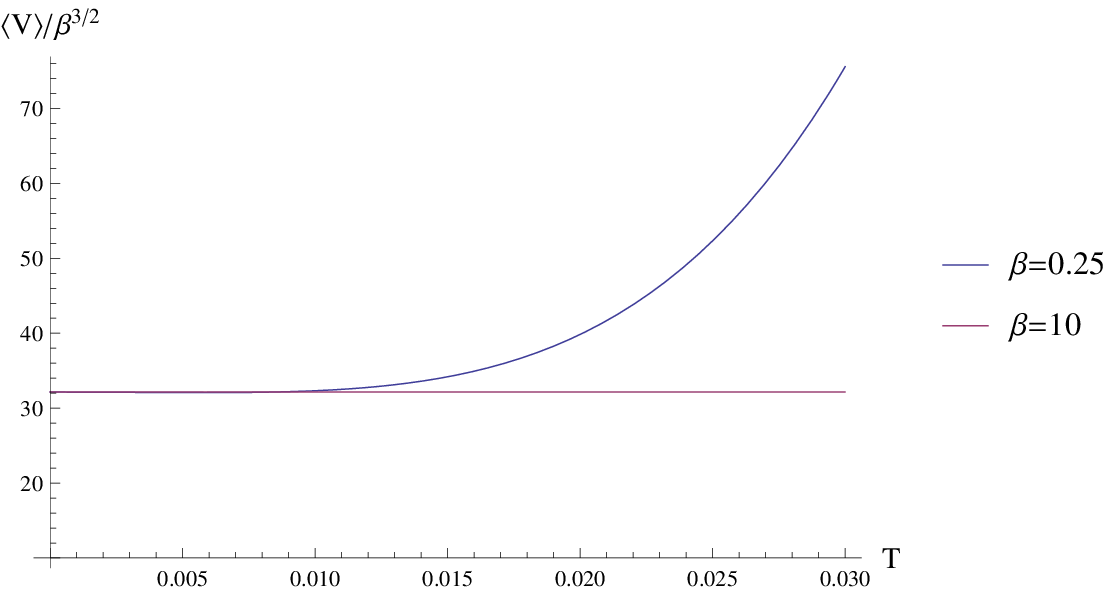}
  \caption{Evolution of the expectation value $\mean{V}$ of the volume operator with an initial eigenvector with eigenvalue $v=32.1396$.}
  \label{EVS25B0}
 \end{figure}
 
 \clearpage

 \begin{figure}[!h]
  \centering
  \includegraphics[width=0.6\textwidth]{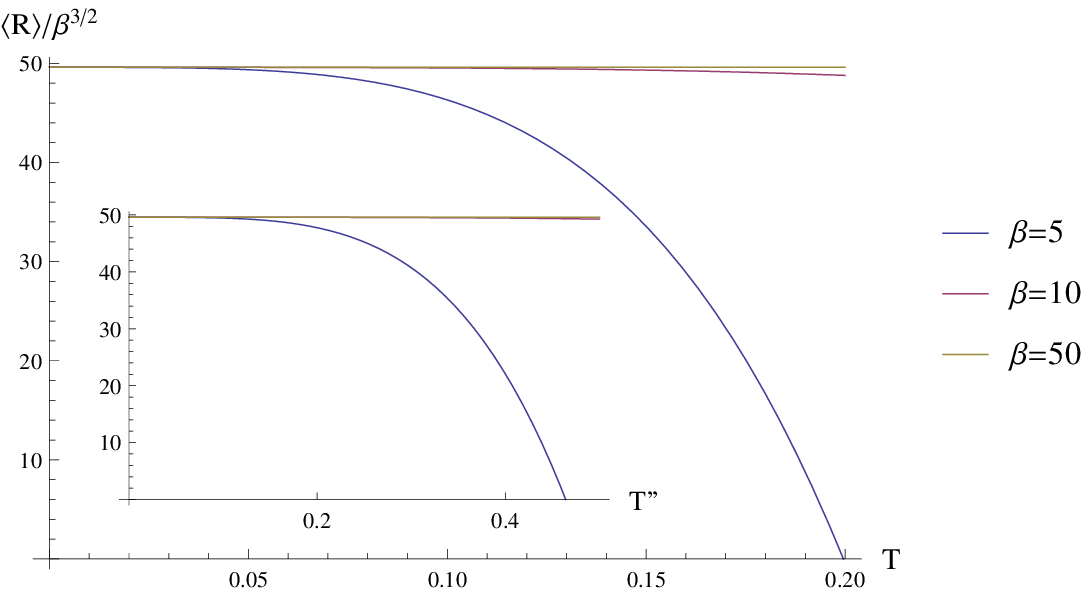}
  \caption{Evolution of the expectation value $\mean{R}$ of the curvature operator with an initial eigenvector with eigenvalue $v=20.1761$.}
  \label{ERS25B}
 \end{figure}
 
 
  \begin{figure}[!h]
  \centering
  \includegraphics[width=0.6\textwidth]{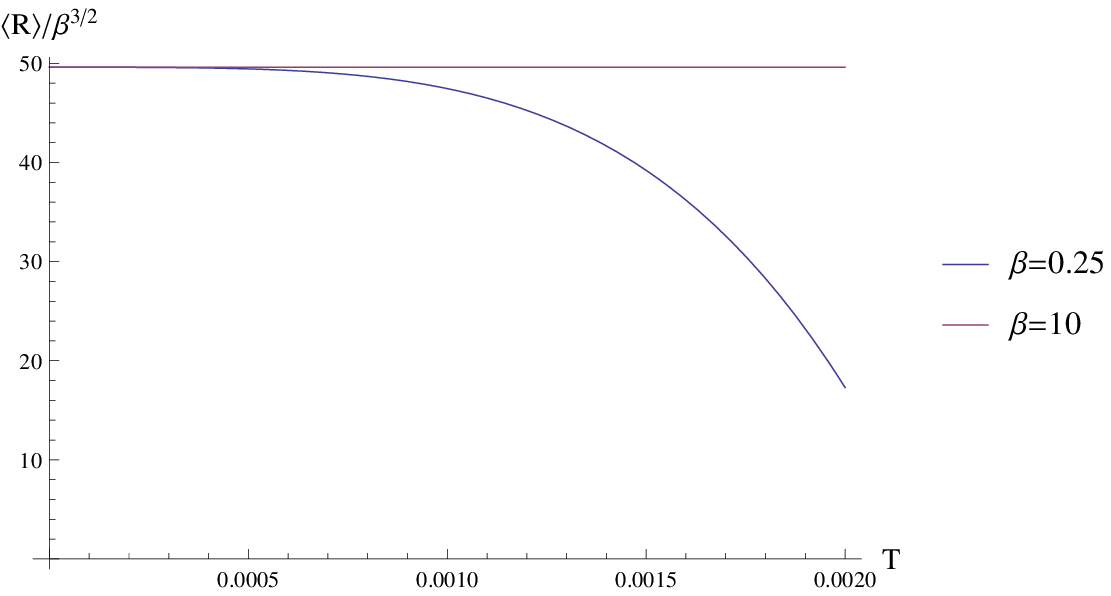}
  \caption{Evolution of the expectation value $\mean{R}$ of the curvature operator with an initial eigenvector with eigenvalue $v=20.1761$.}
  \label{ERS25B0}
 \end{figure}
 
 
\end{itemize}

\begin{figure}[!h]
  \centering
  \includegraphics[width=0.6\textwidth]{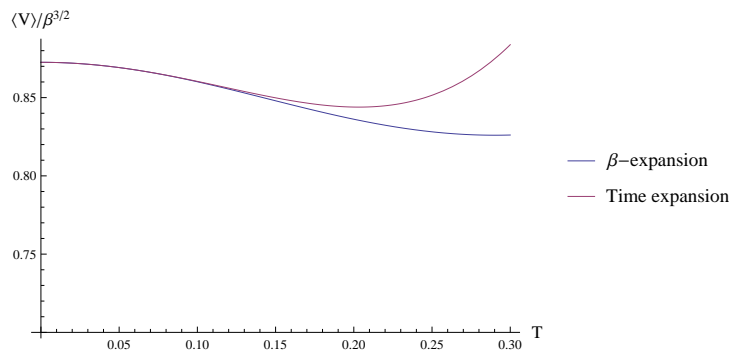}
  \caption{Comparative plot of the evolution of the volume expectation value $\mean{V}$ between the time expansion and the $\beta$-expansion, with an initial eigenvector with eigenvalue $v=0.5730$ and $\beta=50$.}
  \label{Compar}
 \end{figure}

\begin{figure}[!h]
  \centering
  \includegraphics[width=0.6\textwidth]{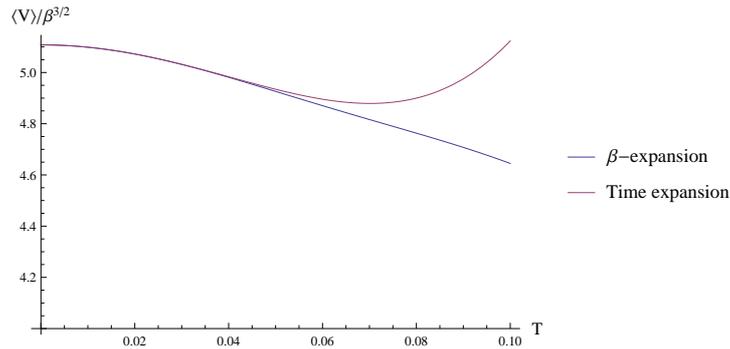}
  \caption{Comparative plot of the evolution of the volume expectation value $\mean{V}$ between the time expansion and the $\beta$-expansion, with an initial eigenvector with eigenvalue $v=5.1078$ and $\beta=100$.}
  \label{Compar1}
 \end{figure}

In figures \ref{Compar} and \ref{Compar1}, we compare the results given by the time expansion and the $\beta$-expansion for $\langle V(T)\rangle$ in a particular volume eigenstate, for $j=2$ and $j=10$ respectively with $\beta=10$. For this value of $\beta$, the $\beta$-expansion presumably provides an accurate description of the dynamics over a longer time interval than the time expansion does. In both figures we observe that around a certain time $T_0$, different in each case, the expectation value given by the time expansion begins to differ significantly from the expectation value given by the $\beta$-expansion. At this time we expect the latter to still be a very close approximation to the exact expectation value.
 
 \clearpage 

\section{Conclusion}

The simulations presented in this article show that the proposed $\beta$-approximation can indeed be applied fully and consistently in a certain sector of the physical Hilbert space in both deparametrized models we considered. Therefore, this perturbation method presents itself as a promising tool in the investigation of the dynamics in LQG models.

A very interesting outcome is the periodic character of the evolution of the expectation values of the volume and curvature operators, in the deparametrized model with a free scalar field. While it is a rather unexpected result, it is clearly reminiscent to the particular physical Hamiltonian present in the model, and the choice of the volume and curvature operators as observables. This indeed suggests the presence of a special relation between the spectral decompositions of the mentioned operators which is yet to be understood.

As a future work, the focus will be on establishing more accurately to which extent one could apply this approximation with respect to the admissible range of the Barbero-Immirzi parameter $\beta$ and the choice of initial states.

%

\begin{center}
 \large{\bf{Acknowledgments}}
\end{center}
This work was supported by the grant of Polish Narodowe Centrum Nauki nr 2011/02/A/ST2/00300. I.M. would like to thank the Jenny and Antti Wihuri Foundation for support.
\clearpage

\appendix

\section{Second-order perturbation theory of a degenerate energy level}\label{app1}

In section \ref{sec:approximation}, we use time-independent perturbation theory to obtain an approximate spectral decomposition of the physical Hamiltonian, treating the Euclidean part of the Hamiltonian as a perturbation over the Lorentzian part. In this case some of the eigenvalues of the unperturbed Hamiltonian are degenerate, and all matrix elements of the perturbation vanish between the degenerate (unperturbed) eigenstates. In such a situation the derivation of the corrections to the eigenstates is not entirely standard, hence we give the full treatment of the perturbative problem up to second order in this appendix.

To conform to the standard notation, in this appendix we write the Hamiltonian\footnote{Precisely speaking, the operator $H$ in eq. \eqref{H0+eV} is the Hamiltonian only in the case of the dust field model, while for the scalar field model the physical Hamiltonian is the square root of an operator of the form \eqref{H0+eV}.} as
\be\label{H0+eV}
H = H_0 + \epsilon V
\ee
where $H_0$ stands for the Lorentzian part and $V$ for the Euclidean part. The perturbation parameter is $\epsilon \equiv 1/(1+\beta^2)$.

Let us denote the eigenvalues and the corresponding eigenstates by $\lambda_n^{(0)}$ and $\kket{\lambda_n^{(0)}}$ for the unperturbed Hamiltonian $H_0$, and $\lambda_n$ and $\kket{\lambda_n}$ for the full Hamiltonian $H$. (For the sake of clarity, we deviate here from the notation of section \ref{sec:approximation}, where the eigenstates and eigenvalues of the unperturbed operator were denoted by $\lambda_n$ and $\kket{\lambda_n}$, and those of the full operator by $\lambda_n'$ and $\kket{\lambda_n'}$.) To determine $\lambda_n$ and $\kket{\lambda_n}$ approximately up to second order in $\epsilon$, we write
\begin{align}
\kket{\lambda_n} &= \kket{\lambda_n^{(0)}} + \epsilon\kket{\lambda_n^{(1)}} + \epsilon^2\kket{\lambda_n^{(2)}} + {\cal O}(\epsilon^3), \\
\lambda_n &= \lambda_n^{(0)} + \epsilon\lambda_n^{(1)} + \epsilon^2\lambda_n^{(2)} + {\cal O}(\epsilon^3).
\end{align}
Inserting these into eq. \eqref{H0+eV}, we obtain
\begin{align}
\bigl(H_0+\epsilon V\bigr)\bigl(\kket{\lambda_n^{(0)}} &+ \epsilon\kket{\lambda_n^{(1)}} + \epsilon^2\kket{\lambda_n^{(2)}} + \dots\bigr) \notag \\
&= \bigl(\lambda_n^{(0)} + \epsilon\lambda_n^{(1)} + \epsilon^2\lambda_n^{(2)} + \dots\bigr)\bigl(\kket{\lambda_n^{(0)}} + \epsilon\kket{\lambda_n^{(1)}} + \epsilon^2\kket{\lambda_n^{(2)}} + \dots\bigr)\label{master-eq}
\end{align}
as the equation from which the corrections to the eigenvalues and eigenstates will be determined.

The derivation of the corrections to the eigenvalues presents no special problems. As is well known, for a degenerate eigenvalue, the first-order corrections are given by the eigenvalues of the matrix of the perturbation in the degenerate subspace. Therefore,
\be\label{lambda(1)}
\lambda_n^{(1)} = 0.
\ee
For the second-order corrections one finds, considering the $\epsilon^2$ terms of eq. \eqref{master-eq},
\be\label{lambda(2)}
\lambda_n^{(2)} = \bbra{\lambda_n^{(0)}}V\kket{\lambda_n^{(1)}} = \sum_k \frac{\bigl|\bbra{\lambda_n^{(0)}}V\kket{\lambda_k^{(0)}}\bigr|^2}{\lambda_n^{(0)}-\lambda_k^{(0)}}.
\ee
In the second equality we used eq. \eqref{psi(1)} to obtain the explicit expression for $\lambda_n^{(2)}$. This step is correct because we are interested in the case where the perturbation $V$ vanishes within the degenerate subspace, implying that the state $V\kket{\lambda_n^{(0)}}$ has no non-vanishing components on unperturbed eigenstates with eigenvalue $\lambda_n^{(0)}$.


Let us then go on to the corrections to the eigenstates. The projections of $\kket{\lambda_n^{(1)}}$ and $\kket{\lambda_n^{(2)}}$ onto unperturbed eigenstates outside the degenerate subspace are easily found by considering the first- and second-order terms of eq. \eqref{master-eq}. We have
\be\label{psi(1)}
\bbrakket{\lambda_k^{(0)}}{\lambda_n^{(1)}} = \frac{\bbra{\lambda_k^{(0)}}V\kket{\lambda_n^{(0)}}}{\lambda_n^{(0)}-\lambda_k^{(0)}} \qquad (\lambda_n^{(0)}\neq\lambda_k^{(0)})
\ee
and, recalling that the first-order correction to the eigenvalue vanishes,
\be
\bbrakket{\lambda_k^{(0)}}{\lambda_n^{(2)}} = \frac{\bbra{\lambda_k^{(0)}}V\kket{\lambda_n^{(1)}}}{\lambda_n^{(0)}-\lambda_k^{(0)}} \qquad (\lambda_n^{(0)}\neq\lambda_k^{(0)}).
\ee
Below we will show that the correction $\kket{\lambda_n^{(1)}}$ has no non-vanishing components on unperturbed eigenstates having eigenvalue $\lambda_n^{(0)}$ -- see eq. \eqref{psi(1)'}. Therefore, using eq. \eqref{psi(1)}, we obtain
\be
\bbrakket{\lambda_k^{(0)}}{\lambda_n^{(2)}} = \sum_l \frac{\bbra{\lambda_k^{(0)}}V\kket{\lambda_l^{(0)}}\bbra{\lambda_l^{(0)}}V\kket{\lambda_n^{(0)}}}{\bigl(\lambda_n^{(0)}-\lambda_k^{(0)}\bigr)\bigl(\lambda_n^{(0)}-\lambda_l^{(0)}\bigr)}\qquad (\lambda_n^{(0)}\neq\lambda_k^{(0)}). \label{psi(2)}
\ee
To find the projections $\bbrakket{\lambda_{n'}^{(0)}}{\lambda_n^{(1)}}$ and $\bbrakket{\lambda_{n'}^{(0)}}{\lambda_n^{(2)}}$, where $\kket{\lambda_{n'}^{(0)}}$ is another unperturbed eigenstate having eigenvalue $\lambda_n^{(0)}$ under the unperturbed Hamiltonian, requires more care. The first-order terms of eq. \eqref{master-eq} do not give any information about $\bbrakket{\lambda_{n'}^{(0)}}{\lambda_n^{(1)}}$; they merely reproduce
\be
\bbra{\lambda_{n'}^{(0)}}V\kket{\lambda_n^{(0)}} = 0
\ee
as a consistency condition for the perturbative expansion. In our case this condition is satisfied irrespectively of the choice of basis in the degenerate subspace. If we turn to the second-order terms of eq. \eqref{master-eq}, we again find no information on $\bbrakket{\lambda_{n'}^{(0)}}{\lambda_n^{(1)}}$, because the first-order correction to the eigenvalue vanishes. Instead, we obtain another consistency condition,
\be
\bbra{\lambda_{n'}^{(0)}} V\kket{\lambda_n^{(1)}} = \sum_k \frac{\bbra{\lambda_{n'}^{(0)}} V\kket{\lambda_k^{(0)}}\bbra{\lambda_{k}^{(0)}} V\kket{\lambda_n^{(0)}}}{\lambda_n^{(0)}-\lambda_k^{(0)}} = 0.
\ee
By a numerical evaluation of the sum, we find that this condition also seems to be satisfied for any choice of basis in the degenerate subspace.

To determine the projections $\bbrakket{\lambda_{n'}^{(0)}}{\lambda_n^{(1)}}$, we must therefore look at the third-order terms in eq. \eqref{master-eq}. We find
\be
\bbra{\lambda_{n'}^{(0)}}V\kket{\lambda_n^{(2)}} = \lambda_n^{(2)}\bbrakket{\lambda_{n'}^{(0)}}{\lambda_n^{(1)}}.
\ee
A computation of the second-order corrrections $\lambda_n^{(2)}$ from eq. \eqref{lambda(2)} confirms that all of them are non-vanishing, at least in the examples considered in section \ref{sec:examples}. Therefore the above equation determines the projections $\bbrakket{\lambda_{n'}^{(0)}}{\lambda_n^{(1)}}$ as
\be
\bbrakket{\lambda_{n'}^{(0)}}{\lambda_n^{(1)}} = \frac{1}{\lambda_n^{(2)}}\bbra{\lambda_{n'}^{(0)}}V\kket{\lambda_n^{(2)}}.
\ee
When the perturbation $V$ is the Euclidean part of the Hamiltonian, the matrix element on the right-hand side actually vanishes. To see this, let us resolve the matrix element in the basis of the unperturbed eigenstates as
\be\label{psi0-V-psi2}
\bbra{\lambda_{n'}^{(0)}}V\kket{\lambda_n^{(2)}} = \sum_k \bbra{\lambda_{n'}^{(0)}}V\kket{\lambda_k^{(0)}}\bbrakket{\lambda_k^{(0)}}{\lambda_n^{(2)}}.
\ee
Here the unperturbed eigenstate $\kket{\lambda_{n'}^{(0)}}$ is a spin-network state based on a single graph, which consists of some number $L$ of special loops attached to a loopless "initial" graph (this is because the Lorentzian part of the Hamiltonian is a graph-preserving operator). The Euclidean part of the Hamiltonian changes the number of loops by one; hence the intermediate states $\kket{\lambda_k^{(0)}}$ entering the sum in eq. \eqref{psi0-V-psi2} have $L-1$ or $L+1$ special loops. On the other hand, by eq. \eqref{psi(2)}, the second-order correction to the state $\kket{\lambda_n^{(0)}}$ is composed of states having $L-2$, $L$ and $L+2$ special loops\footnote{The use of eq. \eqref{psi(2)} in eq. \eqref{psi0-V-psi2} is correct, because we find no degeneracy in the eigenvalues of the Lorentzian part between states based on graphs having a different number of special loops -- with the exception of the eigenvalue zero, which occurs in the dust field model for every graph. However, in this case each eigenstate of the Lorentzian part having eigenvalue zero is also annihilated by the Euclidean part, implying that the matrix element $\bbra{\lambda_{n'}^{(0)}}V\kket{\lambda_k^{(0)}}$ vanishes when $\lambda_{n}^{(0)} = \lambda_k^{(0)} = 0$. Hence the summation index $k$ in eq. \eqref{psi0-V-psi2} always runs only over states whose unperturbed eigenvalue is different from $\lambda_n^{(0)}$.}. Therefore the scalar product $\bbrakket{\lambda_k^{(0)}}{\lambda_n^{(2)}}$ on the right-hand side of eq. \eqref{psi0-V-psi2} is always zero, and we conclude that
\be\label{psi(1)'}
\bbrakket{\lambda_{n'}^{(0)}}{\lambda_n^{(1)}} = 0.
\ee
It still remains to determine the components of the second-order correction $\kket{\lambda_n^{(2)}}$ within the degenerate subspace. The projection of $\kket{\lambda_n^{(2)}}$ on the uncorrected eigenstate $\kket{\lambda_n^{(0)}}$ can be found by requiring that the corrected eigenstate $\kket{\lambda_n^{(0)}} + \epsilon\kket{\lambda_n^{(1)}} + \epsilon^2\kket{\lambda_n^{(2)}}$ is normalized up to second order in $\epsilon$. In this way we find
\be\label{psi(2)'}
\bbrakket{\lambda_n^{(0)}}{\lambda_n^{(2)}} = -\frac{1}{2}\bbrakket{\lambda_n^{(1)}}{\lambda_n^{(1)}} = \sum_k \frac{\bigl|\bbra{\lambda_n^{(0)}}V\kket{\lambda_k^{(0)}}\bigr|^2}{\bigl(\lambda_n^{(0)}-\lambda_k^{(0)}\bigr)^2}.
\ee
The projection $\bbrakket{\lambda_{n'}^{(0)}}{\lambda_n^{(2)}}$, where $\kket{\lambda_{n'}^{(0)}}$ is another unperturbed eigenstate sharing the degenerate eigenvalue $\lambda_n^{(0)}$, is not uniquely determined by any normalization or orthogonality conditions. In our application to the physical Hamiltonian, this projection is also not determined by the equations obtained from \eqref{master-eq}, at least up to sixth order in $\epsilon$, apparently reflecting the fact that the degeneracy present in the eigenvalues of the Lorentzian part of the Hamiltonian is not removed by the Euclidean part at second order of perturbation theory. We resolve this situation by choosing
\be\label{psi(2)''}
\bbrakket{\lambda_{n'}^{(0)}}{\lambda_n^{(2)}} = 0,
\ee
this choice being the simplest, and consistent with the normalization and orthogonality of the corrected eigenstates up to second order in $\epsilon$. This completes the derivation of the corrected eigenvalues and eigenstates up to second order, the solution being given by eqs. \eqref{lambda(1)}, \eqref{lambda(2)}, \eqref{psi(1)}, \eqref{psi(2)}, \eqref{psi(1)'}, \eqref{psi(2)'} and \eqref{psi(2)''}.


\section{Perturbative expansion of expectation values}\label{app2}

In equation \eqref{Gen.Exp.}, the general and compact expression for the time-dependent expectation value of an operator $A$ in the scalar field model was given. Here we display explicitly the different contributions to this expectation value, organized order by order in the perturbation. In general, using the notation of the previous Appendix, we have
\begin{align}
 \nonumber \mean{A(T)} =\prod \limits_{x\in \Sigma}\ \sum \limits_{\substack{i,j=1\\ \lambda_i^\prime \geq 0 \\ \lambda_j^\prime \geq 0}}^{d_x} &\exp\biggl[ \frac{-i}{\hbar} T \sqrt{\frac{1+\beta^2}{16\pi G\beta^2}} \left(\sqrt{\lambda_i^{(0)} + \epsilon\lambda_i^{(1)} + \epsilon^2\lambda_i^{(2)}}-\sqrt{\lambda_j^{(0)} + \epsilon\lambda_j^{(1)} + \epsilon^2\lambda_j^{(2)}}\right)\biggr]\\ \nonumber & \left[\langle\Psi_0|\lambda_j^{(0)}\rangle \bra{\lambda_j^{(0)}}A \ket{\lambda_i^{(0)}}\langle\lambda_i^{(0)}|\Psi_0\rangle + \right.\\ \nonumber
 & + \epsilon \left( \langle\Psi_0|\lambda_j^{(1)}\rangle \bra{\lambda_j^{(0)}}A \ket{\lambda_i^{(0)}}\langle\lambda_i^{(0)}|\Psi_0\rangle + \langle\Psi_0|\lambda_j^{(0)}\rangle \bra{\lambda_j^{(1)}}A \ket{\lambda_i^{(0)}}\langle\lambda_i^{(0)}|\Psi_0\rangle + \right.\\ \nonumber & \qquad \left. + \langle\Psi_0|\lambda_j^{(0)}\rangle \bra{\lambda_j^{(0)}}A \ket{\lambda_i^{(1)}}\langle\lambda_i^{(0)}|\Psi_0\rangle + \langle\Psi_0|\lambda_j^{(0)}\rangle \bra{\lambda_j^{(0)}}A \ket{\lambda_i^{(0)}}\langle\lambda_i^{(1)}|\Psi_0\rangle \right) + \\ \nonumber
 & + \epsilon^2 \left( \langle\Psi_0|\lambda_j^{(2)}\rangle \bra{\lambda_j^{(0)}}A \ket{\lambda_i^{(0)}}\langle\lambda_i^{(0)}|\Psi_0\rangle + \langle\Psi_0|\lambda_j^{(0)}\rangle \bra{\lambda_j^{(2)}}A \ket{\lambda_i^{(0)}}\langle\lambda_i^{(0)}|\Psi_0\rangle + \right.\\ \nonumber & \qquad \left. + \langle\Psi_0|\lambda_j^{(0)}\rangle \bra{\lambda_j^{(0)}}A \ket{\lambda_i^{(2)}}\langle\lambda_i^{(0)}|\Psi_0\rangle + \langle\Psi_0|\lambda_j^{(0)}\rangle \bra{\lambda_j^{(0)}}A \ket{\lambda_i^{(0)}}\langle\lambda_i^{(2)}|\Psi_0\rangle + \right.\\ \nonumber
 & \qquad + \left. \langle\Psi_0|\lambda_j^{(1)}\rangle \bra{\lambda_j^{(1)}}A \ket{\lambda_i^{(0)}}\langle\lambda_i^{(0)}|\Psi_0\rangle + \langle\Psi_0|\lambda_j^{(0)}\rangle \bra{\lambda_j^{(1)}}A \ket{\lambda_i^{(1)}}\langle\lambda_i^{(0)}|\Psi_0\rangle + \right.\\ \nonumber
 & \qquad + \left. \langle\Psi_0|\lambda_j^{(1)}\rangle \bra{\lambda_j^{(0)}}A \ket{\lambda_i^{(1)}}\langle\lambda_i^{(0)}|\Psi_0\rangle + \langle\Psi_0|\lambda_j^{(0)}\rangle \bra{\lambda_j^{(1)}}A \ket{\lambda_i^{(0)}}\langle\lambda_i^{(1)}|\Psi_0\rangle+ \right.\\ 
 & \qquad + \left. \left. \langle\Psi_0|\lambda_j^{(1)}\rangle \bra{\lambda_j^{(0)}}A \ket{\lambda_i^{(0)}}\langle\lambda_i^{(1)}|\Psi_0\rangle + \langle\Psi_0|\lambda_j^{(0)}\rangle \bra{\lambda_j^{(0)}}A \ket{\lambda_i^{(1)}}\langle\lambda_i^{(1)}|\Psi_0\rangle \right) \right].
\end{align}

However, a large number of terms in this general expression actually vanish in the case that is of interest to us. When the operator $A$ is graph-preserving, the initial state $\ket{\Psi_0}$ is based on a single graph, and the unperturbed Hamiltonian and the perturbation are respectively the Lorentzian and the Euclidean operators, the above expression for the expectation value simplifies to the following:
\begin{align}
 \mean{A(T)} =\prod \limits_{x\in \Sigma}\ \sum \limits_{\substack{i,j=1\\ \lambda_i^\prime \geq 0 \\ \lambda_j^\prime \geq 0}}^{d_x} &\exp\biggl[ \frac{-i}{\hbar} T \sqrt{\frac{1+\beta^2}{16\pi G\beta^2}} \left(\sqrt{\lambda_i^{(0)} + \epsilon^2\lambda_i^{(2)}}-\sqrt{\lambda_j^{(0)} + \epsilon^2\lambda_j^{(2)}}\right)\biggr]\\ \nonumber & \left[\langle\Psi_0|\lambda_j^{(0)}\rangle \bra{\lambda_j^{(0)}}A \ket{\lambda_i^{(0)}}\langle\lambda_i^{(0)}|\Psi_0\rangle + \right.\\ \nonumber
 & + \epsilon^2 \left( \langle\Psi_0|\lambda_j^{(2)}\rangle \bra{\lambda_j^{(0)}}A \ket{\lambda_i^{(0)}}\langle\lambda_i^{(0)}|\Psi_0\rangle + \langle\Psi_0|\lambda_j^{(0)}\rangle \bra{\lambda_j^{(2)}}A \ket{\lambda_i^{(0)}}\langle\lambda_i^{(0)}|\Psi_0\rangle + \right.\\ \nonumber & \qquad \left. + \langle\Psi_0|\lambda_j^{(0)}\rangle \bra{\lambda_j^{(0)}}A \ket{\lambda_i^{(2)}}\langle\lambda_i^{(0)}|\Psi_0\rangle + \langle\Psi_0|\lambda_j^{(0)}\rangle \bra{\lambda_j^{(0)}}A \ket{\lambda_i^{(0)}}\langle\lambda_i^{(2)}|\Psi_0\rangle + \right.\\
 & \nonumber \qquad + \left. \langle\Psi_0|\lambda_j^{(0)}\rangle \bra{\lambda_j^{(1)}}A \ket{\lambda_i^{(1)}}\langle\lambda_i^{(0)}|\Psi_0\rangle + \left. \langle\Psi_0|\lambda_j^{(1)}\rangle \bra{\lambda_j^{(0)}}A \ket{\lambda_i^{(0)}}\langle\lambda_i^{(1)}|\Psi_0\rangle \right) \right].
\end{align}

\clearpage

\bibliographystyle{plainnat}

\end{document}